\documentclass[a4paper] {article}
\usepackage{graphics,color,epsfig}
\def\linadj#1{\normalbaselines
	\multiply\lineskip#1 \divide\lineskip100
 	\multiply\baselineskip#1 \divide\baselineskip100
	\multiply\lineskiplimit#1 \divide\lineskiplimit100 }

\begin{document}








\title{\bf Effect of finite chemical potential on QGP-Hadron phase transition in a statistical model of fireball formation.}

\author{ R. Ramanathan, Agam K. Jha$^+$, \\  K. K. Gupta$^*$ and S. S. Singh}

\maketitle
\begin{center}

 \Large  Department of Physics, University of Delhi, Delhi - 110007, INDIA  \\ \Large$^+$Department of  physics, Sri Venkateswara College, University of Delhi,Delhi,  $^*$Department of Physics, Ramjas College, University of Delhi, Delhi - 110007, INDIA

\end{center}
\linadj{200}

\begin{abstract}
\large We study the effect of finite chemical potential for the QGP constituentsin the Ramanathan et al.statistical model (Phys.Rev.C70, 027903,2004). While the earlier computations using this model with vanishing chemical potentials indicated a weakly first order phase transition for the system in the vicinity of $170~MeV$ (Pramana, 68, 757, 2007), the introduction of finite values for the chemical potentials of the constituents makes the transition a smooth roll over of the phases, while allowing fireball formation with radius of a few 'fermi' to take place. This seems to be in conformity with the latest consensus on the nature of the QGP-Hadron phase transition.\\


PACS number(s): 25.75.Ld, 12.38.MH, 21.65.+f

\end {abstract}

\vfill
\eject

\maketitle
 
\section{Introduction}

\large  The formation of QGP droplet(fireball) is one of the most exciting possibility in ultra relativistic heavy ion collisions(URHIC) [1].The physics of such an event is very complicated and to extract meaningful results from a rigorous use of QCD appropriate to this physical system is almost intractable though heroic efforts at lattice estimation of the problem has been going on for quite some time[2]. One way out is to replicate the approximation schemes which have served as theoretical tools in understanding equally complicated atomic and nuclear systems in atomic and nuclear physics in the context of QGP droplet formation.This approach lays no claim to rigour or ab-initio '¡Èunderstanding'¡É of the phenomenon but lays the  framework on which more rigorous structures may be built depending on the phenomenological success of the model as and when testable data emerges from ongoing experiments.
 
The nucleation process is driven by statistical fluctuations  which produce the QGP droplets in a hadronic phase,the size of the fluctuations being determined by the critical free energy difference between the two phases.The Kapusta et-al model (3) uses the liquid drop model expansion for this,as given by

\begin{eqnarray} 
\Delta F = \frac {4\pi}{3} R^{3} [P_ {had}(T,\mu_{B}) - P_{q,g}(T, \mu_{B})] \nonumber \\
  + 4\pi R^{2} \sigma +\tau_{crit} T~ln \biggl [ 1 + (\frac {4 \pi}{3})R^{3}s_{q,g} \biggl].                                                                     \end{eqnarray}   

The first term represents the volume contribution,the second term is the surface contribution where $\sigma$ is the surface tension,and the last term is the so called shape contribution.The shape contribution is  an entropy term on account of fluctuations in droplet shape which we may ignore in the lowest order approximation.The critical radius $R_{c}$ can be obtained by minimising (1) withrespect to the droplet radius R,which in the Linde approximation[6] is,

 \begin{equation}
 R_{c} = \frac {2\sigma}{\Delta p}~ or~                                             \sigma = \frac {3\Delta F_{c}}{4\pi R_{c}^{2}}
\end{equation}

\section{The statistical model}
In the approximation scheme of the Ramanathan et.al[4 ] the relativistic density of states for the quarks and gluons is constructed adapting the procedures of the Thomas-Fermi construction of the electronic density of states for complex atoms and the Bethe density of states[6] for nucleons in complex nuclei as templates .The expression for the density of states for the quarks and gluons(q,g) in this model is:

\begin{equation}\label{3.13}
\rho_{q, g} (k) = (\nu / \pi^2) \lbrace (-V_{conf}(k))^{2}(-\frac{dV_{conf}(k)}{dk}) \rbrace_{q, g}, 
\end{equation}

where k is the relativistic four-momentum of the quarks and the gluons,$\nu$ is the volume of the fire ball taken to be a constant in the first approximation and V is a suitable confining potential relevant to the current quarks and gluons in the QGP [4] given by:

 \begin{equation}\label{3.18}
V_{\mbox{eff}}(k) = (1/2k)\gamma_{g,q} ~ g^{2} (k) T^{2} - m_{0}^{2} / 2k~.
\end{equation}
                    
where m is the mass of the quark which we take as zero for the up and down quarks and 150Mev.for the strange quarks.The g(k) is the QCD running coupling constant given by

\begin{equation}\label{3.17}
g^2(k) =(4/3)(12\pi/27)\lbrace 1/ \ln(1+k^{2}/\Lambda^{2}) \rbrace ~,
\end{equation}

where $\Lambda$ is the QCD scale taken to be $150~ MeV$.

The model has a natural low energy cut off at:

\begin{equation}\label{3.19}
k_{min}=V(k_{min})~or~ k_{min}=(\gamma_{g,q}N^{\frac{1}{3}} T^{2} \Lambda^2 / 2)^{1/4},
\end{equation}

with $N=[(4/3 )(12 \pi / 27)]^{3}$.

The free energy of the respective cases i(quarks,gluons,interface etc.) for Fermions and Bosons(upper sign or lower sign) can be computed using the following expression:

\begin{equation}\label{3.20}
F_i = \mp T g_i \int dk \rho_i (k) \ln (1 \pm e^{-(\sqrt{m_{i}^2 + k^2}) /T})~,
\end{equation}

With the surface free-energy given by a modified Weyl[7] expression:
\begin{equation}\label{3.21}
F_{interface} = \gamma T \int dk \rho_{weyl} (k) \delta (k-T)~,
\end{equation}

where the hydrodynamical flow parameter for the surface is:

\begin{equation}\label{3.22}
\gamma = \sqrt{2}\times \sqrt{(1/\gamma_{g})^{2} + (1 / \gamma_{q})^{2}},
\end{equation}

For the pion which for simplicity represents the hadronic medium in which the fire ball resides ,the free energy is:

\begin{equation}\label{3.25}
F_{\pi} = (3T/2\pi^2 )\nu \int_0^{\infty} k^2 dk \ln (1 - e^{-\sqrt{m_{\pi}^2 + k^2} / T})~.
\end{equation}

With these ingredients we can compute the free-energy change with respect to both the droplet radius and temperature to get a physical picture of the fire ball formation,the nucleation rate governing the droplet formation ,the nature of the phase transition etc.This can be done over a whole range of flow-parameter values [4],We exhibit only the two most promising scenarios in fig.$1$ and fig.$2$. Further  investigating the properties of the corresponding free energies as in [4],it was found that the QGP-hadron phase transition is a weakly first order one.
\begin{figure}
\begin{center}
\epsfig{figure=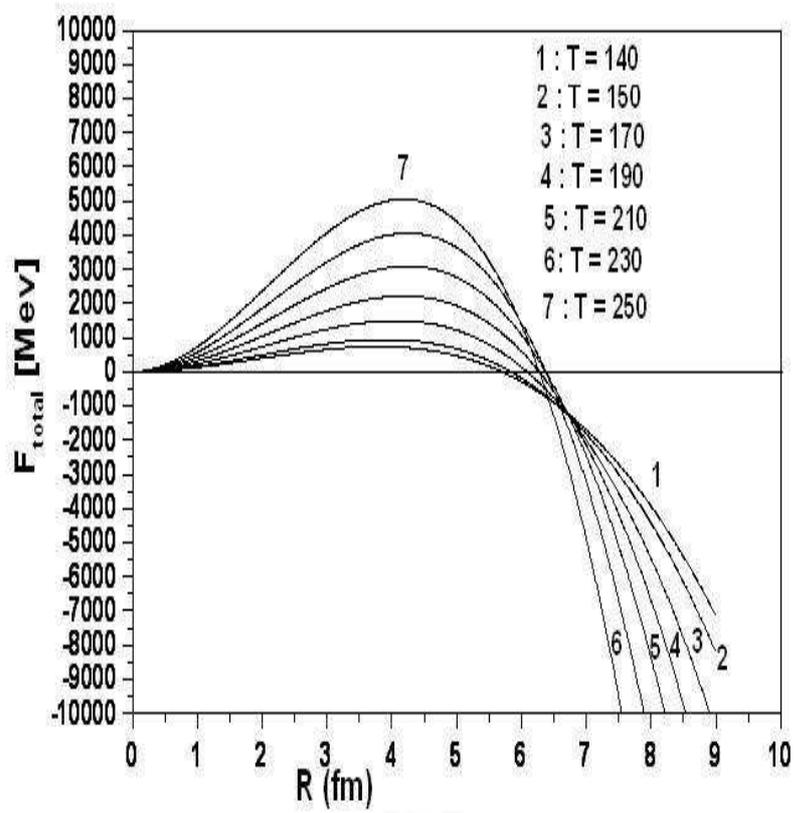,height=5.5in,width=4.5in}
\label{fig4}
\caption{\large $F_{total}$ at $\gamma_{g} = 6\gamma_{q}$, $ \gamma_{q} = 1/6 $ for various temperatures.}
\end{center}
\end{figure}

\begin{figure}
\begin{center}
\epsfig{figure=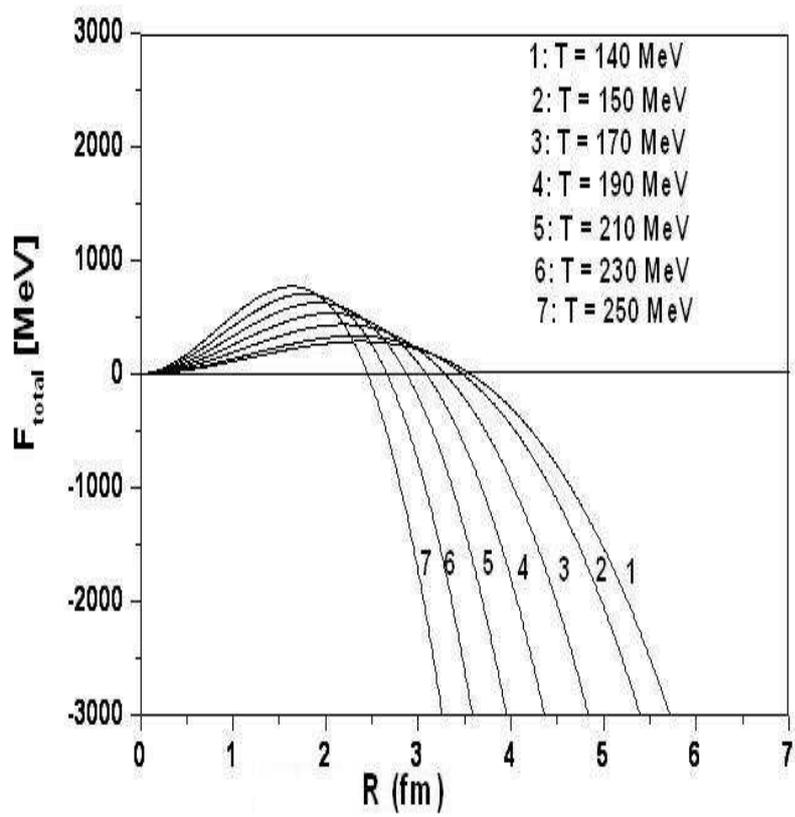,height=5.5in,width=4.5in}
\label{fig5}
\caption{\large $F_{total}$ at $\gamma_{g} = 8\gamma_{q}$ , $ \gamma_{q} = 1/6 $ for various temperatures.}
\end{center}
\end{figure}
\vfill
\eject

\section{Effect of finite chemical potential on the phase transition}

All our previous calculations were based on the assumption of vanishing chemical potential for the quarks,which is not a realistic one. Therefore, we consider the changes in our result by introducing finite chemical potentials into our formulation. This is achieved by ,as is done in statistical mechanics ,by changing the exponents in  (7), from $-\frac{E}{T}$ to $\frac{\mu-E}{T}$, where '$\mu$'is the chemical potential. Following the typical values of the chemical potentials for the quarks and gluons in the literature [8], we use at $300~MeV$ and $400~MeV$ for the different parameter values of the model.The result of our computations are displayed in figs.$3$ to $14$.

\begin{figure}
\epsfig{figure=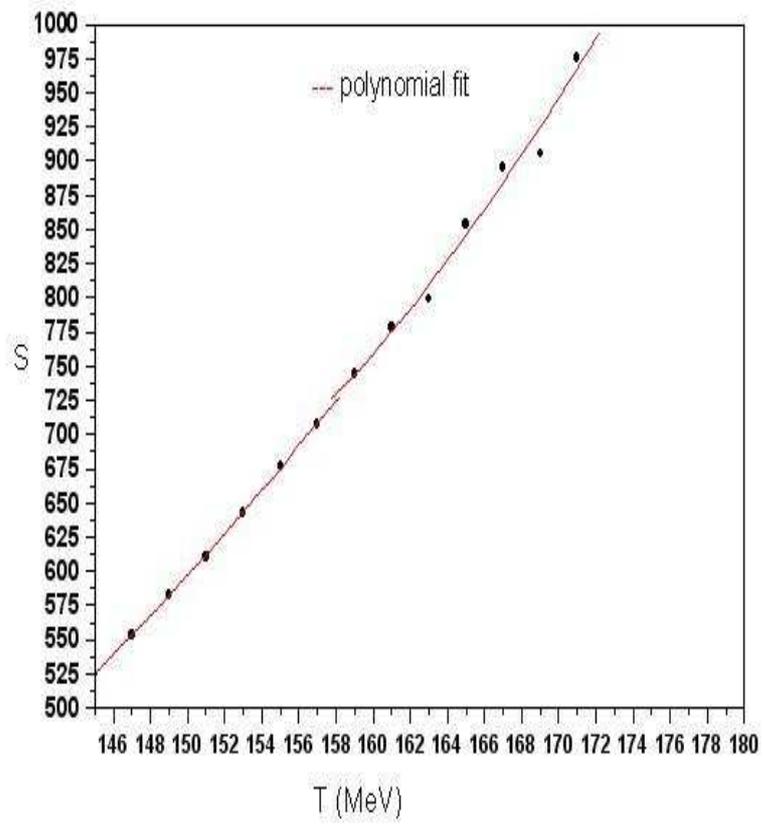,height=5.5in,width=4.5in}
\label{st61.eps}
\caption{\large  Variation of $S$ with temperature T at $\gamma_{g} = 6\gamma_{q}$ , $ \gamma_{q} = 1/6 $.}
\end{figure}
 \begin{figure}
\epsfig{figure=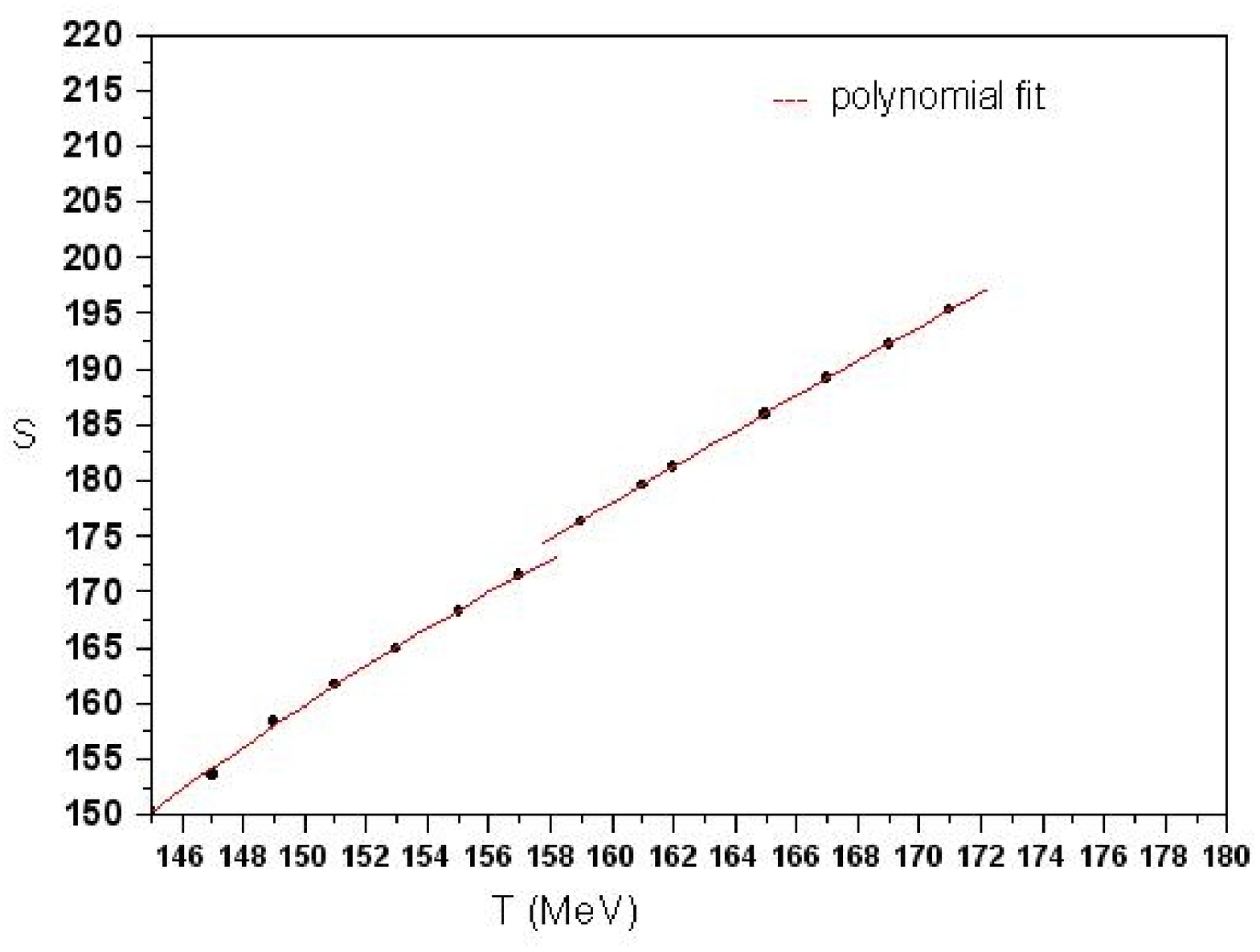,height=5.5in,width=4.5in}
\label{st81.eps}
\caption{\large  Variation of $S$ with temperature T at $\gamma_{g} = 8\gamma_{q}$ , $ \gamma_{q} = 1/6 $.}
\end{figure}

 \begin{figure}
\epsfig{figure=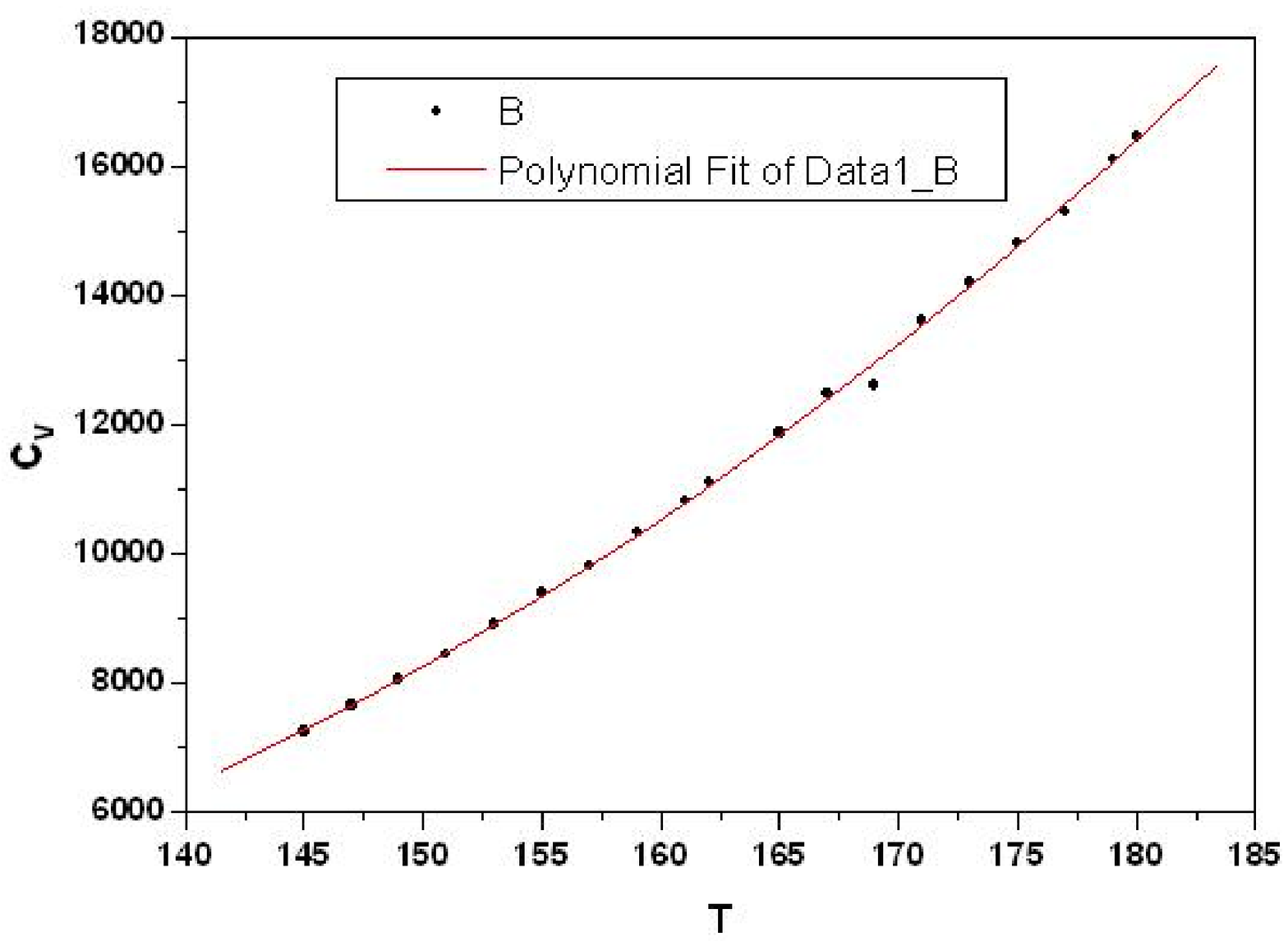,height=5.5in,width=4.5in}
\label{cv6.eps}
\caption{\large Variation of specific heat $C_{V}$ with temperature T at $\gamma_{g} = 6\gamma_{q}$ , $ \gamma_{q} = 1/6 $.}
\end{figure}

\begin{figure}
\epsfig{figure=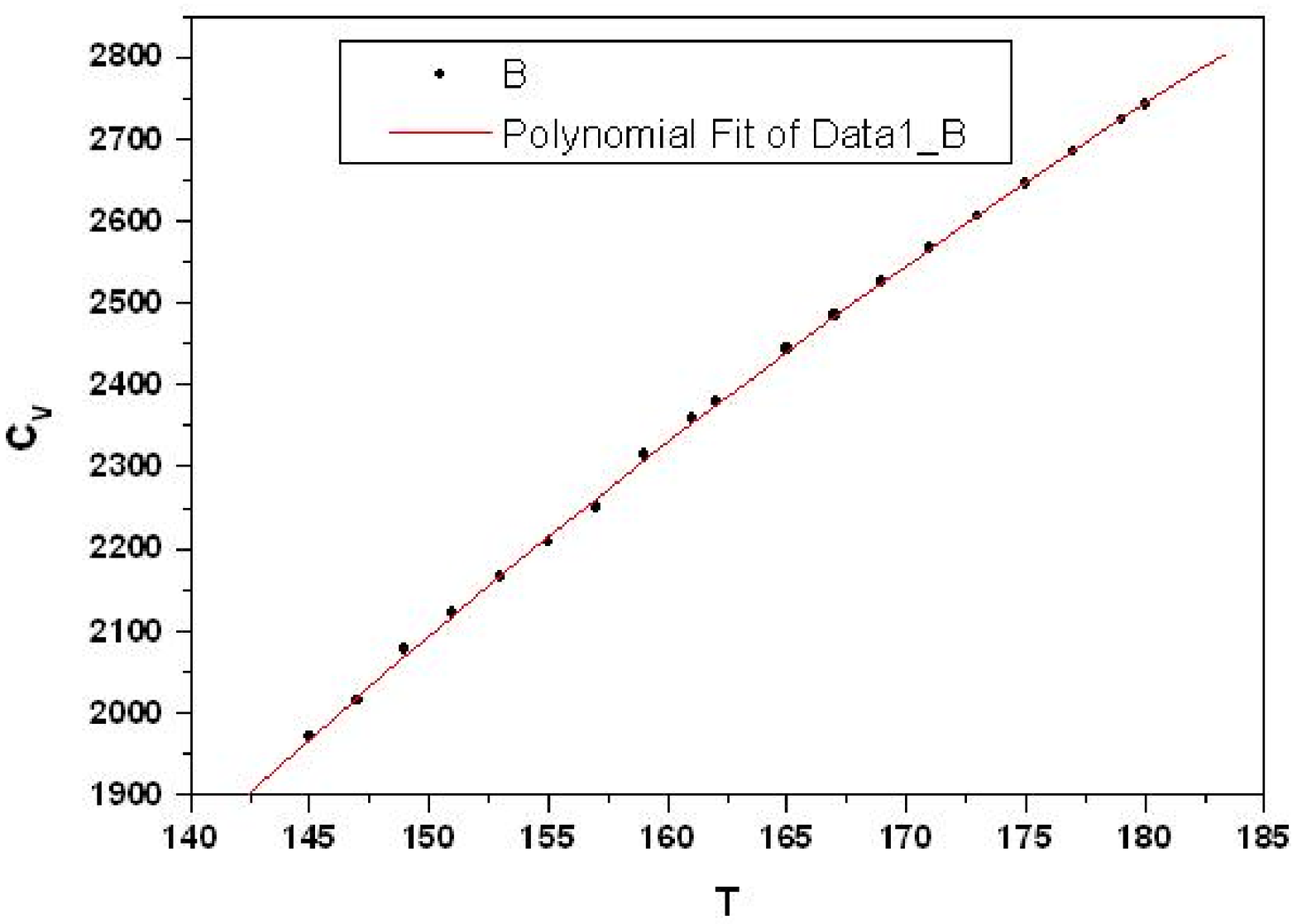,height=5.5in,width=4.5in}
\label{cv8.eps}
\caption{\large Variation of specific heat $C_{V}$ with temperature T at $\gamma_{g} = 8\gamma_{q}$ , $ \gamma_{q} = 1/6 $.}
\end{figure}

\begin{figure}
\epsfig{figure=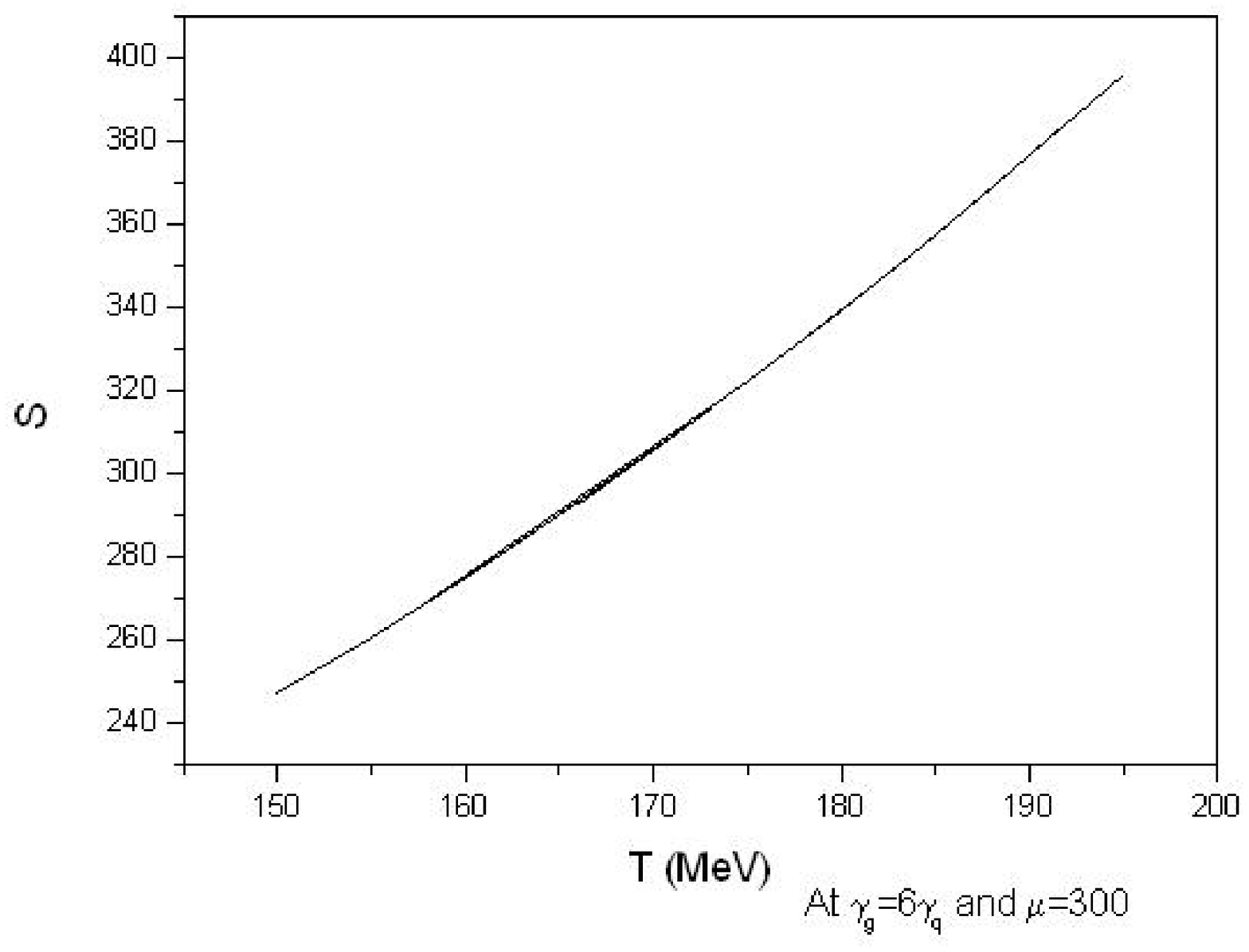,height=5.5in,width=4.5in}
\label{mu3006.eps}
\caption{\large  Variation of $S$ with temperature T at $\gamma_{g} = 6\gamma_{q}$ , $ \gamma_{q} = 1/6$  at $\mu=300$ .}
\end{figure}
 \begin{figure}
\epsfig{figure=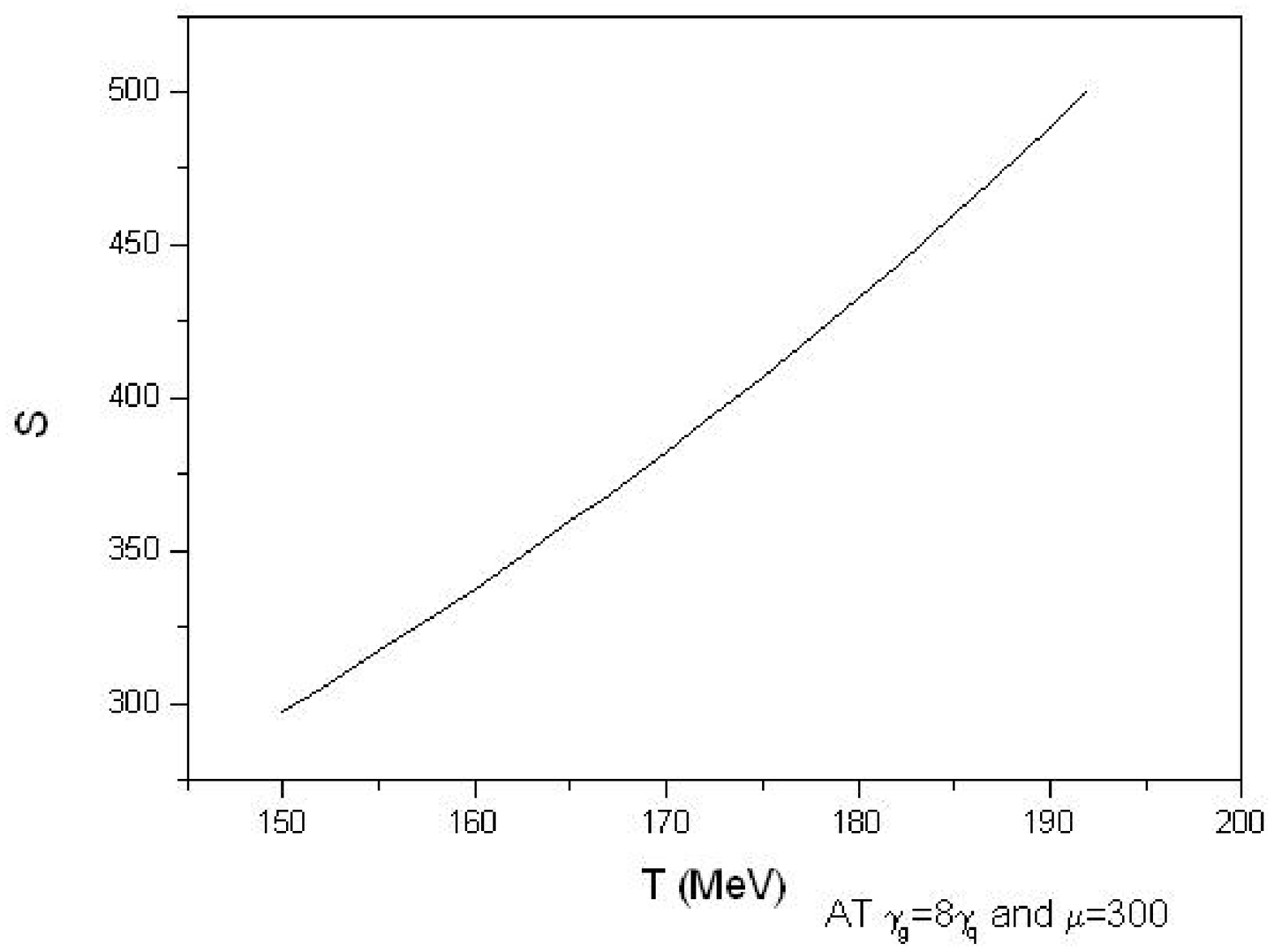,height=5.5in,width=4.5in}
\label{mu3008.eps}
\caption{\large  Variation of $S$ with temperature T at $\gamma_{g} = 8\gamma_{q}$ , $ \gamma_{q} = 1/6 $ at $\mu=300$ .}
\end{figure}

\begin{figure}
\epsfig{figure=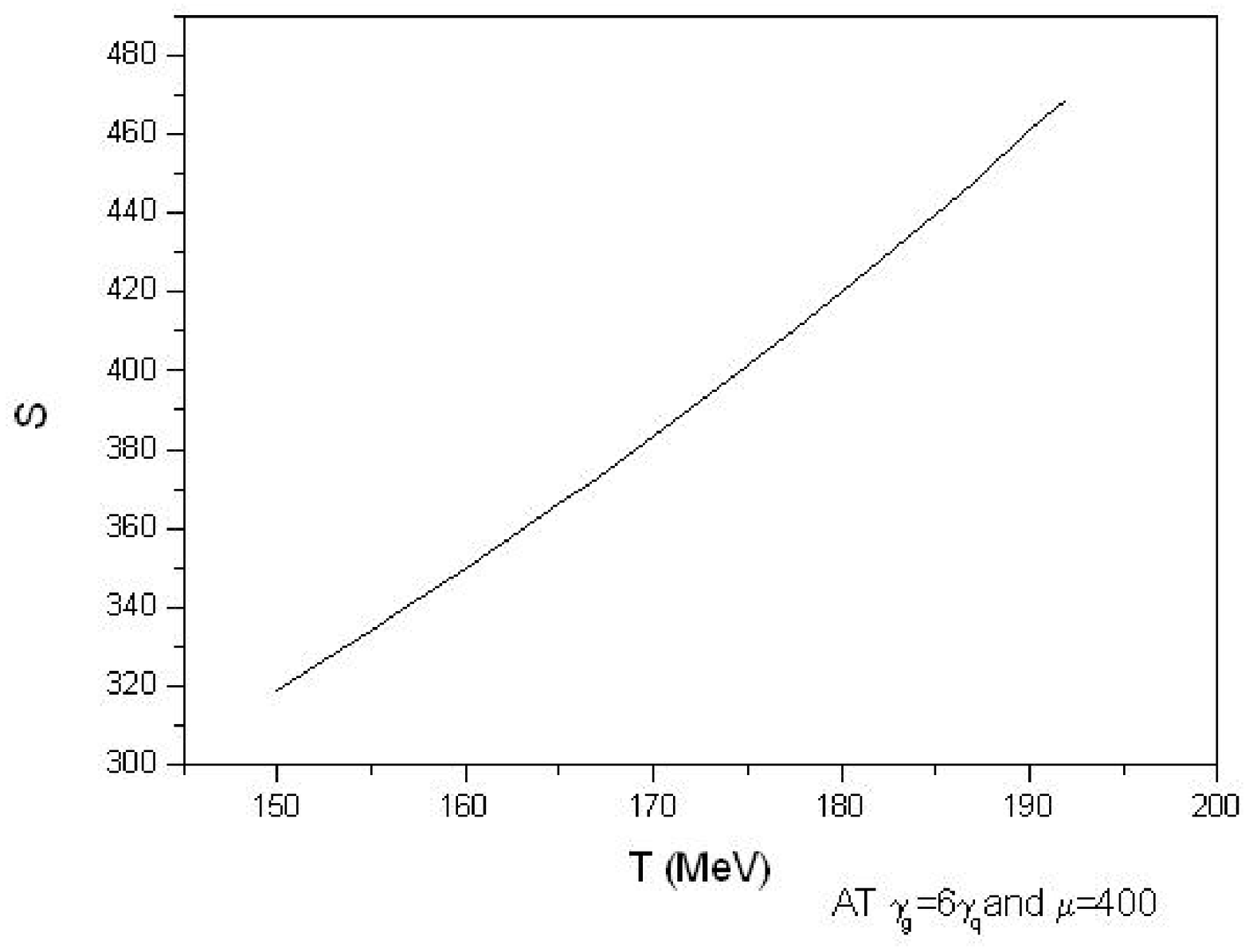,height=5.5in,width=4.5in}
\label{mu4006.eps}
\caption{\large  Variation of $S$ with temperature T at $\gamma_{g} = 6\gamma_{q}$ , $ \gamma_{q} = 1/6$  at $\mu=400$ .}
\end{figure}
 \begin{figure}
\epsfig{figure=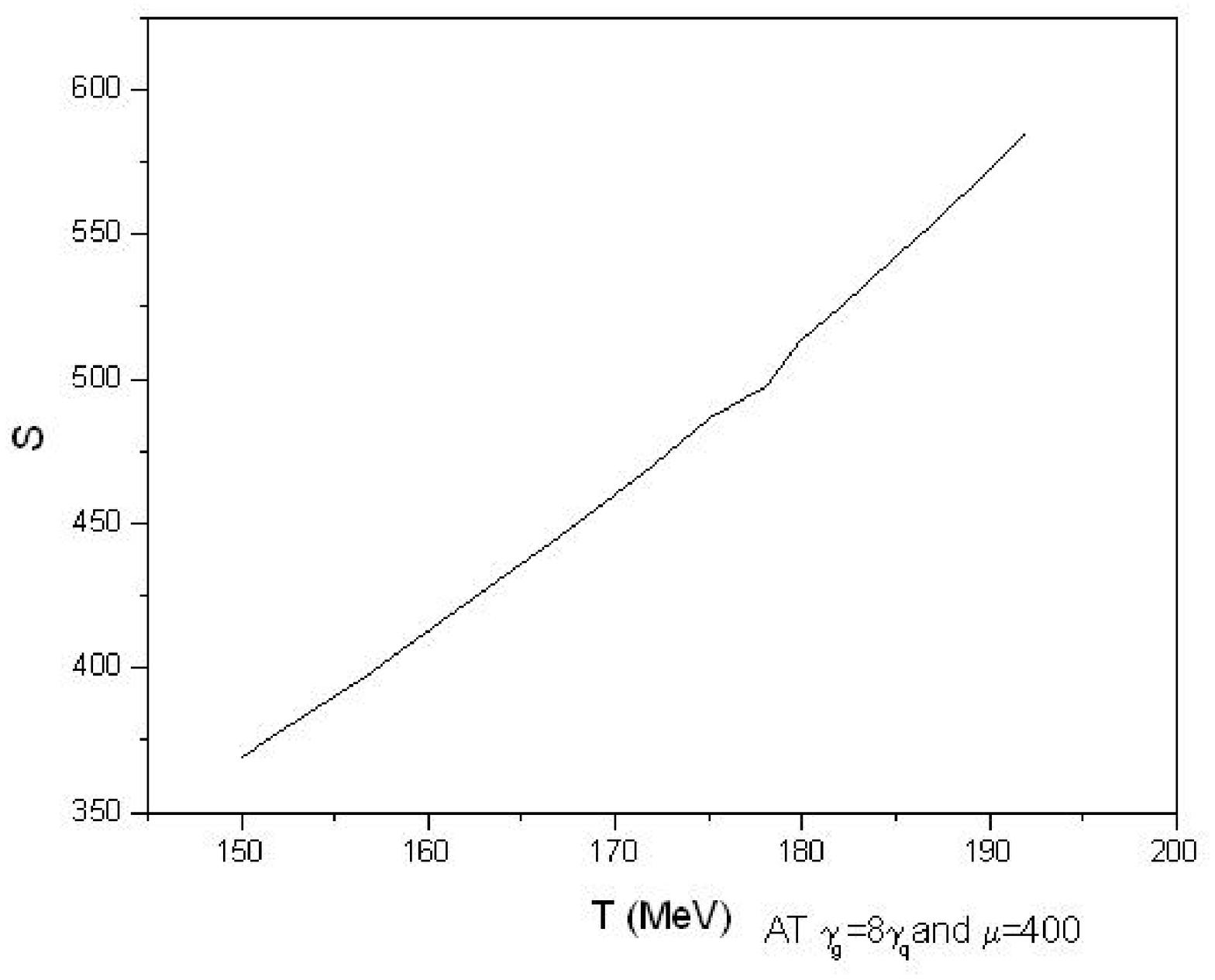,height=5.5in,width=4.5in}
\label{mu4008.eps}
\caption{\large  Variation of $S$ with temperature T at $\gamma_{g} = 8\gamma_{q}$ , $ \gamma_{q} = 1/6 $ at $\mu=300$ .}
\end{figure}

\begin{figure}
\epsfig{figure=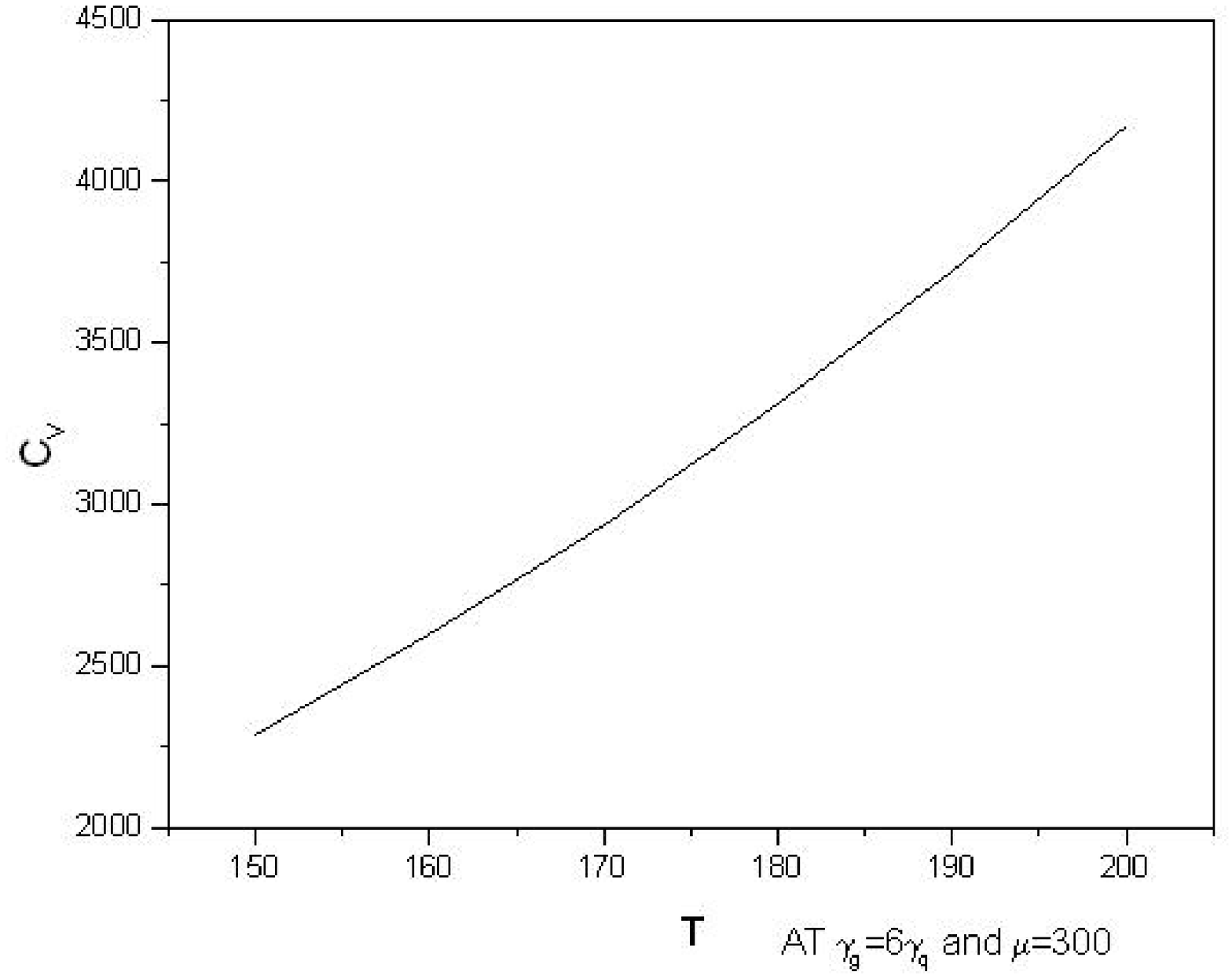,height=5.5in,width=4.5in}
\label{cvm3006.eps}
\caption{\large  Variation of $C_{V}$ with temperature T at $\gamma_{g} = 6\gamma_{q}$ , $ \gamma_{q} = 1/6$  at $\mu=300$ .}
\end{figure}
 \begin{figure}
\epsfig{figure=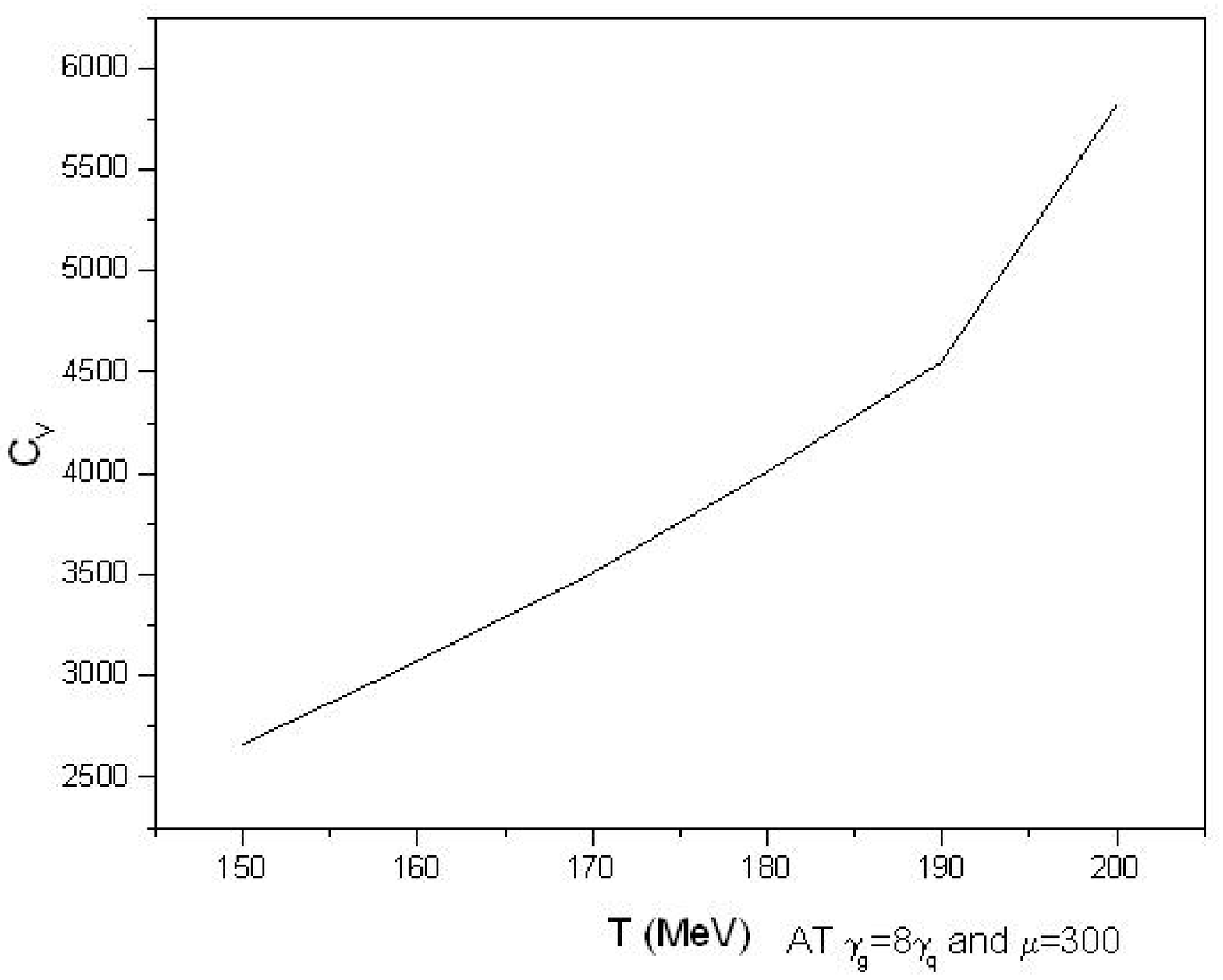,height=5.5in,width=4.5in}
\label{cvm3008.eps}
\caption{\large  Variation of $C_{V}$ with temperature T at $\gamma_{g} = 8\gamma_{q}$ , $ \gamma_{q} = 1/6 $ at $\mu=300$ .}
\end{figure}

\begin{figure}
\epsfig{figure=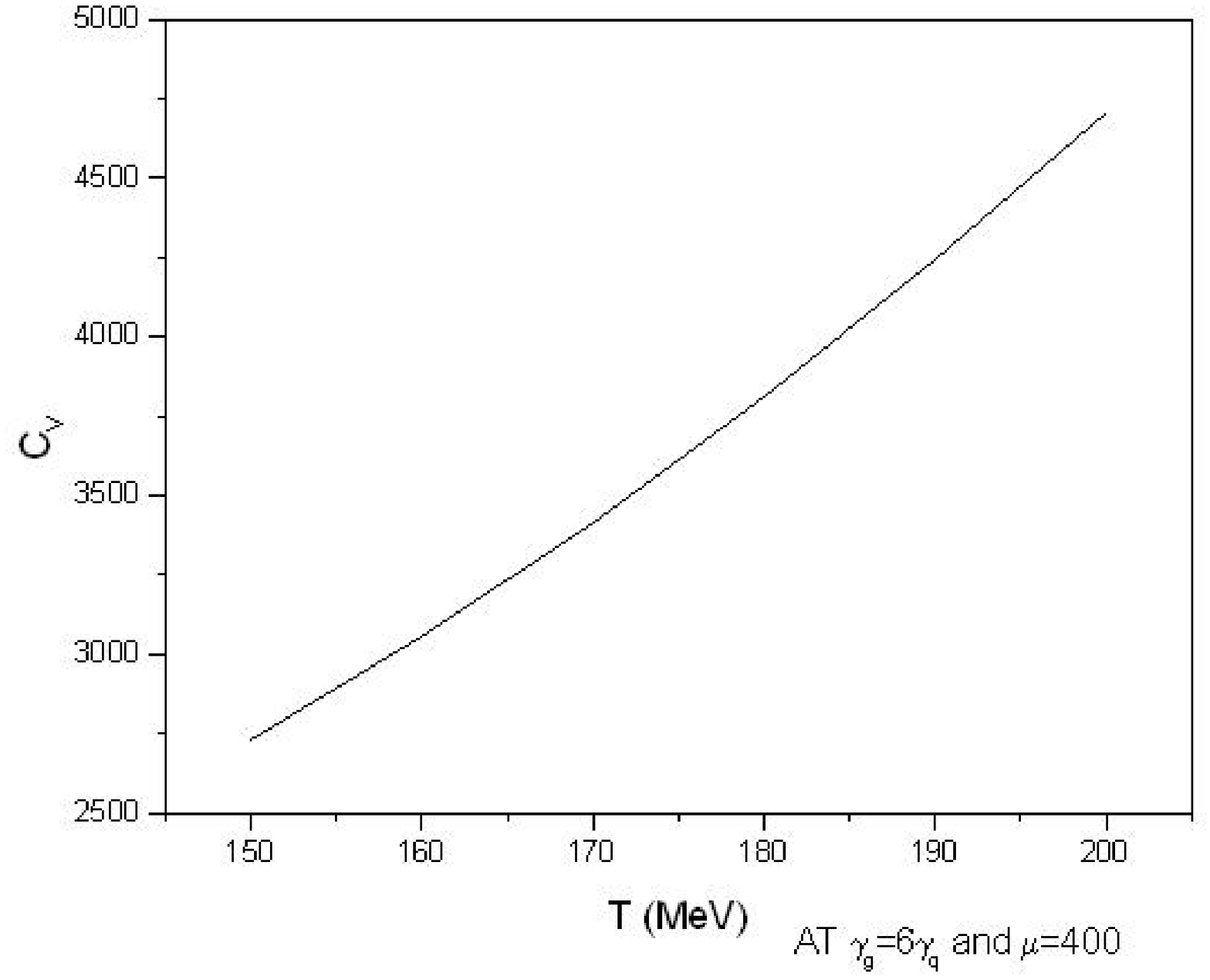,height=5.5in,width=4.5in}
\label{cvm4006.eps}
\caption{\large  Variation of $C_{V}$ with temperature T at $\gamma_{g} = 6\gamma_{q}$ , $ \gamma_{q} = 1/6$  at $\mu=400$ .}
\end{figure}
 \begin{figure}
\epsfig{figure=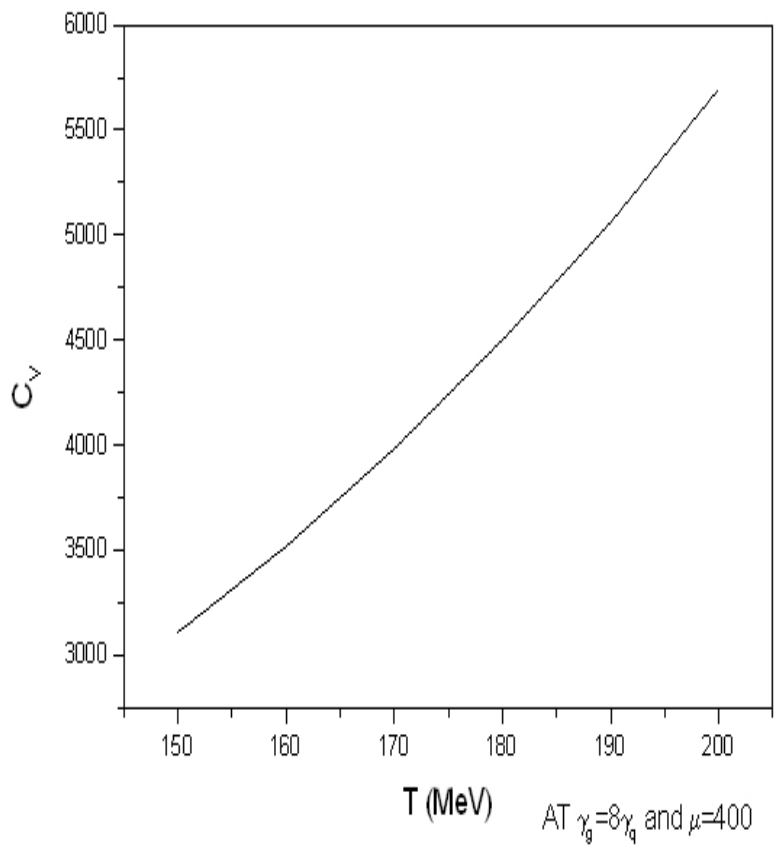,height=5.5in,width=4.5in}
\label{cvm4008.eps}
\caption{\large  Variation of $C_{V}$ with temperature T at $\gamma_{g} = 8\gamma_{q}$ , $ \gamma_{q} = 1/6 $ at $\mu=400$ .}
\end{figure}
\vfill
\eject

As could be seen from the above graphs the effect of finite chemical potential seems to make the phase transition a smooth roll over from the hadron phase to the QGP phase rather than a weakly first order transition predicted by the model for the unrealistic zero chemical potential case [4].This result is in consonance with the present expectations from lattice simulations [8].

{\bf References :}
\begin{enumerate}
\item{F. Karsch, E. Laermann, A. Peikert, Ch. Schmidt and S. Stickan, Nucl. Phys. B (proc. Suppl.) 94, 411 (2001).}
 \item{T. Renk, R. Schneider, and W. Weise, Phys. Rev. C 66, 014902 (2002).}
\item{L. P. Csernai, J. I. Kapusta, R. Venugopalan, and E. Osnes, Phys. Rev. D  67, 045003 (2003); J. I. Kapusta, R. Venugopalan, and  A. P. Vischer,        Phys. Rev. C 51, 901 (1995); L. P. Csernai and J. I. Kapusta, Phys. Rev. D 46, 1379 (1992). P. Shukla and A. K. Mohanty,  Phys. Rev. C 64, 054910 (2001).}

\item{R. Ramanathan, Y. K. Mathur,  K. K. Gupta, and Agam K. Jha Phys.Rev.C70,027903 (2004); ; R. Ramanathan, ,  K. K. Gupta, Agam K. Jha, and S.S.Singh, Pramana 68,757 (2007).}
\item{B.D.Malhotra and R.Ramanathan,Phys.Lett.A108:153,(1985).}
\item{E. Fermi, Z. Phys. 48, 73 (1928); L. H. Thomas, Proc. Cambridge Philos. Soc. 23, 542 (1927); H. A. Bethe, Rev. Mod. Phys. 9, 69 (1937).}
 \item{H. Weyl, Nachr. Akad. Wiss Gottingen 110 (1911).}
 \item{G.Neergaard and J. Madsen, Phys. Rev. D 62, 034005 (2000).}
	
\end{enumerate}

\end{document}